\documentclass[aps,prl,preprintnumbers,groupedaddress,showpacs,twocolumn,floatfix]{revtex4}

\usepackage{graphicx}

\begin{document}

\newcommand{\bfm}[1]{\mbox{\boldmath$#1$}}
\newcommand{\bff}[1]{\mbox{\scriptsize\boldmath${#1}$}}

\title{Quantum Hall Effect in Quantum Electrodynamics}

\author{Alexander A.~Penin}
\email{apenin@phys.ualberta.ca}
\affiliation{Department of Physics, University of Alberta,
Edmonton, AB T6G 2J1, Canada}
\affiliation{
Institute for Nuclear Research of Russian Academy of Sciences,
117312 Moscow, Russia}


\begin{abstract}
We consider the quantum Hall effect  in  quantum electrodynamics and
find a deviation from the quantum mechanical prediction for  the
Hall conductivity due to radiative antiscreening  of  electric
charge in an external  magnetic field. A weak  dependence of the
universal von Klitzing constant on the magnetic field strength,
which can possibly be observed in a dedicated experiment, is
predicted.
\end{abstract}


\maketitle


The quantum Hall effect (QHE) \cite{KDP,TSG} is a remarkable
phenomenon remaining in the focus of  experimental and  theoretical
research over the last three decades. The study of the QHE led to
development of new fundamental  physical concepts \cite{Lau2,Hal2}.
At the same time the QHE  plays a crucial role in metrology and
determination of fundamental constants  \cite{MohTay}.

In the QHE, in sharp contrast to the prediction of classical
electrodynamics, the conductivity  of the two-dimensional electron
system in  a strong transverse magnetic field at low temperature has
plateaus as a function of the magnetic field strength. At these
plateaus the conductivity is given by integer or  specific
fractional multiples of $R_K^{-1}$, where $R_K$ is a universal
parameter known as the von Klitzing constant. A simple quantum
mechanical consideration of the noninteracting electron gas relates
it to the fine structure constant \footnote{Throughout the paper, if
it is not explicitly stated otherwise, we adopt the system of units
used in particle physics, where $\hbar=c=1$ and
$\alpha=e^2/(4\pi)$.}
\begin{equation}
 R_K^{-1}=2\alpha\,.
\label{rk}
\end{equation}
A remarkable property of a two dimensional electron system in
magnetic field is that this na\"ive result is stable against all
kinds of perturbations which do not result in a qualitative change
of the Landau spectrum. This has been  proven first in
Ref.~\cite{Lau1} (see also Ref.~\cite{Hal1}) by an elegant use of
gauge invariance. Later, a relation  of the Hall conductivity to the
topological invariants of the adiabatic ground state space has been
established \cite{TKNN,Sim,AvrSei}. Much work has been done to find
a possible deviation from Eq.~(\ref{rk}) (see {\it e.g.}
\cite{Yen}). However, leaving aside  finite temperature and edge
effects, no universal corrections have been found and Eq.~(\ref{rk})
is currently considered to be exact \cite{MohTay}. This would
distinguish the quantum Hall conductance as one of a very few
characteristics of many-particle interacting quantum systems exactly
predicted by theory. On the other hand the exact relation~(\ref{rk})
would allow for determination of the fine structure constant with
{\it a~priori}  zero theoretical uncertainty.

The purpose of this paper is to show that in quantum electrodynamics
(QED) quantum field effects lead to deviation from the  quantum
mechanical prediction for the  Hall conductance. The physics behind
this phenomenon is in a modification  of the  electromagnetic
coupling of electrons due to  vacuum polarization by highly virtual
electron-positron pairs in a  strong magnetic field, which can
roughly be described as radiative antiscreening of the  electric
charge. The main result of the paper, which is the
leading order QED correction to Eq.~(\ref{rk}), is given by
Eq.~(\ref{res}).

\begin{figure}[t]
\begin{center}
\begin{picture}(0,0)
\includegraphics[width=2.5cm]{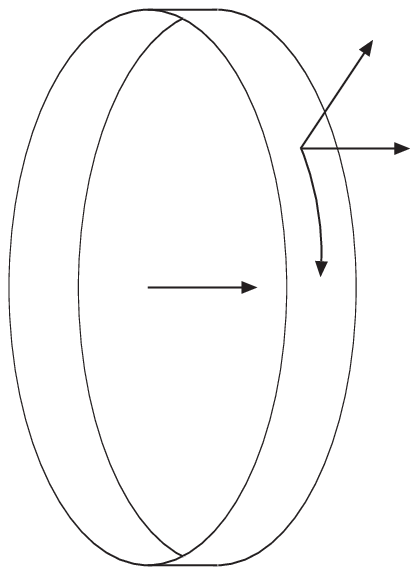}
\end{picture}
\begin{picture}(70,110)
\put(75,69){\makebox(0,0)[r]{$\bfm E$}}
\put(72,95){\makebox(0,0)[r]{$\bfm B$}}
\put(55,45){\makebox(0,0)[r]{$\bfm I$}}
\put(30,41){\makebox(0,0)[r]{$\Phi$}}
\end{picture}
\end{center}

\caption{\small{Geometry of the Hall current. The size of the loop is much larger than
any other scale of the problem and the magnetic field is homogeneous near the surface of the ribbon.} \label{fig1}}
\end{figure}

Following Ref.~\cite{Lau1} we consider the Hall current $I$ around
an asymptotically large loop of a two-dimensional ribbon subject to
a time-independent locally homogeneous magnetic field $B$ and an
electric field $E$.  The spatial vectors  $\bfm{I}$, $\bfm{B}$, and
$\bfm{E}$  are orthogonal to each other and the magnetic field is
normal to the ribbon surface, see Fig.~\ref{fig1}.  For the  futher
analysis it is convenient to introduce an auxiliary magnetic flux
$\Phi$ through the loop.  The  Hall conductivity  $R_H^{-1}$ is
defined by the equation  $I=R_H^{-1}V$ where $V$ is the potential
drop across the ribbon. In QHE it is given by $R_H^{-1}=\nu
R_K^{-1}$ where the filling factor $\nu$ can be either integer
\cite{KDP} or fractional \cite{TSG}. We focus on the integer QHE
since the case of fractional $\nu$ can be understood as  the integer
QHE for  fractionally charged quasiparticles \cite{Lau2}.

The QHE  is a collective phenomena in  condensed matter and the
analysis of quantum field effects in such a system is not
straightforward. To get a systematic description of interacting
electrons in  QED we use the  nonrelativistic effective theory
approach \cite{CasLep}.  Let us briefly outline it. The core idea of
the method is to disentangle the contributions of excitations
corresponding to  widely separated dynamical scales. In the absence
of interaction with the medium, the dynamics of an electron in a
magnetic field  $B$ is characterized by three scales: the hard scale
of the electron mass $m$, the soft scale of the  cyclotron momentum
$\sqrt{eB}$, and the ultrasoft scale of the cyclotron energy $eB/m$.
If the parameter
\begin{equation}
\beta =\frac{eB}{m^2} \label{beta}
\end{equation}
is small, the above scales are widely separated and the effective
field theory method is applicable. Interaction with the medium
results in appearance of additional soft and ultrasoft scales.
However, the specific nature of these scales is not important until
the above hierarchy is violated. The electrons in the ground state
of the Landau spectrum are nonrelativistic and have soft momentum
and ultrasoft energy. The hard  and soft excitations  could only
appear as virtual states and the dynamics of the real electrons is
determined  by an effective Schr\"odinger equation and the multipole
interaction with the ultrasoft  photons \cite{PinSot1}. The
corresponding effective Hamiltonian is of the following form
\begin{equation}
{\cal H}=e^*A_0-\frac{\bfm D^2}{2m^*}+\delta{\cal H}\,, \label{ham}
\end{equation}
where  $A_0$ is the potential of  the electric field $E$,
$\bfm{D}$ is the spatial covariant derivative, $e^*$ $(m^*)$
stands for the effective  charge (mass) of the electron, and
$\delta{\cal H}$ represents  the radiative and relativistic
corrections as well as  the interaction with the medium. The entire
contribution of the hard and soft excitations is encoded in the
parameters of the Hamiltonian, which can be systematically computed
in QED as a series in $\alpha$ and $\beta$ or, in general, a ratio
of the scales present in the problem. The quantum Hall conductivity
is known to be independent of $m^*$  and $\delta{\cal H}$
\cite{Lau1,TKNN}. The ultrasoft contribution represents the effect
of retardation and cannot be reduced to a variation of the
Hamiltonian. Nevertheless, the arguments of Refs.~\cite{Lau1,TKNN}
hold and the ultrasoft contribution to $R_H^{-1}$ vanishes, which
can be checked by an explicit calculation (see {\it e.g.}
Ref.~\cite{KniPen}).

Thus the only source of the corrections to Eq.~(\ref{rk}) is
electron coupling to the external fields. This coupling is modified
by vacuum polarization through creation of hard virtual
electron-positron pairs. In the absence of a  magnetic field this
effect is reabsorbed by the on-shell renormalization of  the
physical electron charge $e$.  For a nonvanishing magnetic field the
vacuum polarization cannot  be ``renormalized out'' and the
effective charge does differ from $e$. Since the magnetic field $B$
explicitly breaks down the Lorentz invariance, the effective charges
are in general different for different external fields. For the
calculation of the Hall conductivity we need  beside $e^*$ another
effective charge $e^\prime$, which parametrizes the coupling of the
electrons to the vector potential of the auxiliary magnetic flux  in
the covariant derivative ${\bfm D}={\bfm \partial}-ie^\prime{\bfm
A}^\Phi+\ldots$.

The effective charges are  determined by the   behavior of the
vacuum polarization tensor $\Pi_{\mu\nu}(q)$ at small four-momentum
transfer $q$. By using the integral representation of
Refs.~\cite{Sch,Adl} it is straightforward to derive the leading
variation of the polarization tensor due to the magnetic field in
the limit $q\to 0$, which reads
\begin{eqnarray}
\lefteqn{\delta\Pi_{\mu\nu}(q)=-\frac{\alpha}{\pi}
\beta^2\frac{1}{45}\bigg[ 2\left(g_{\mu\nu}q^2-q_\mu q_\nu\right)}
\nonumber\\
&&-7\left(g_{\mu\nu}q^2-q_\mu q_\nu\right)_\parallel
+4\left(g_{\mu\nu}q^2-q_\mu q_\nu\right)_\perp\bigg].
\label{delpi}
\end{eqnarray}
The correction to the polarization tensor is transverse  because of
the gauge invariance. At the same time the Lorentz invariance  is
broken and Eq.~({\ref{delpi}}) includes the transverse projectors in
the ``parallel''   $(q_0,\bfm{q}_\parallel)$ and ``orthogonal''
$(\bfm{q}_\perp)$ two-dimensional subspaces of the whole
four-dimensional Minkowskian momentum space $(q_0,\bfm{q})$.  Here
$\bfm{q}_\parallel$ and $\bfm{q}_\perp$ components correspond to the
spatial momentum parallel and orthogonal to the magnetic field,
respectively. The polarization tensor  determines both the correction to the 
local coupling of electrons to the electromagnetic potential and
the correction to the photon propagator, which is non-trivial since the 
external magnetic field changes the photon dispersion law \cite{Adl}.

\begin{figure}[t]
\begin{center}
\begin{picture}(0,0)
\includegraphics[width=8cm]{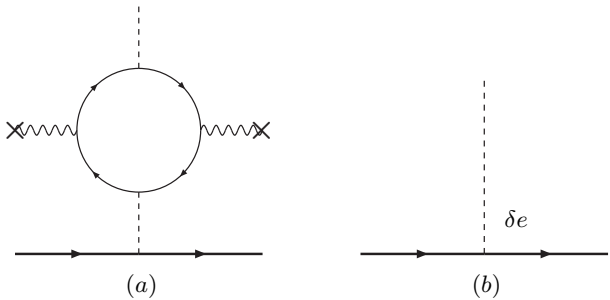}
\end{picture}
\begin{picture}(220,110)
\put(194,15){\makebox(0,0)[r]{$\delta e$}}
\put(54,-15){\makebox(0,0)[rb]{$(a)$}}
\put(184,-15){\makebox(0,0)[rb]{$(b)$}}
\end{picture}
\end{center}

\caption{\small{Feynman diagrams in full QED $(a)$ and in the nonrelativistic
effective theory $(b)$ representing the antiscreening of  the electric charge in
the external magnetic field. The  arrow lines correspond to the free electron propagators.
The bold arrow lines correspond to the electron propagating in the external magnetic field.
The dashed lines represent the electric potential, the crossed wavy lines represent the
external magnetic field, and $\delta e=e^*-e$.} \label{fig2}}
\end{figure}

Let us now consider the effective charge $e^*$, which parametrizes
the interaction of the electron to the homogeneous electric field.
The first two terms in square brackets of Eq.~({\ref{delpi}}) result
in a modification of the static Coulomb potential  between two
pointlike charges \cite{LosSko}
\begin{equation}
V(\bfm{r})=\frac{\alpha}{r}\left[1+\frac{\alpha}{\pi} \beta^2\left(\frac{2}{45}-\frac{7}{90}\sin^2 \theta\right)\right],
\label{pot}
\end{equation}
where $\theta$ is the angle between $\bfm{B}$ and  $\bfm{r}$, {\it
i.e.} the Coulomb interaction in the presence of the magnetic field
becomes anisotropic.  Taking an infinite uniformly charged plane  as
a source  of  $E$ and using the  potential~({\ref{delpi}}) for the
electron interaction with the charge density, one gets the following
result
\begin{equation}
e^*=e\left[1+\frac{1}{45}\frac{\alpha}{\pi} \beta^2\right]\,
\label{estar}
\end{equation}
where the only nonvanishing contribution is due to the first Lorentz covariant
term of Eq.~(\ref{delpi}). The angular dependent term in
Eq.~(\ref{pot}) represents the correction to the Coulomb photon propagator 
and its contribution to  Eq.~(\ref{estar}) vanishes. Thus the  vacuum
polarization in the magnetic field enhances the electron coupling to the
electric field which generates the Hall current. Graphically the effect is
represented by the Feynman diagrams in Fig.~\ref{fig2}.

Similar effect occurs in the case of the effective charge $e^\prime$. 
The vector potential of the auxiliary magnetic flux has only  $\bfm{A}_\perp$
component and its momentum has only  $\bfm{q}_\parallel$ component.
Thus  only the first term of Eq.~({\ref{delpi}}) contributes to the
corresponding coupling and  one gets  $e^\prime
=e\left[1+\alpha\beta^2/(45\pi)\right]$, {\it
i.e.}  $e^\prime=e^*$. Note that $e^\prime$ is
exactly the parameter which appears in the quantization condition
for the auxiliary magnetic flux through the contour of the Hall
current. Hence in the presence of the magnetic field $B$ the
``effective'' flux quantum becomes   $\Phi_0^\prime=2\pi/e^\prime$
or
\begin{equation}
(\Phi_0^\prime)^{-1}=\frac{e}{2\pi}\left[1+\frac{1}{45}\frac{\alpha}{\pi} \beta^2\right].
\label{phistar}
\end{equation}
Now we are in a position to derive the correction to the quantum
mechanical result for the Hall conductivity. In general, the Hall 
current is given by the integral of the current density over the 
ribbon cross section
\begin{equation}
I=\int \left[j_I({\bfm r})+\delta j_I({\bfm r})\right]{\rm d}r_E{\rm
d}r_B\,, \label{cur}
\end{equation}
where  ${\bfm r}=(r_I,r_E,r_B)$ is a vector with the
components parallel to the  ${\bfm I}$,  ${\bfm E}$, and
${\bfm B}$, respectively. A single electron contribution to the 
unperturbed current density can be written as follows
\begin{equation}
j_I({\bfm r})=-i\frac{e}{m}\phi^*({\bfm r})D_I\phi({\bfm r})\,, \label{jdef}
\end{equation}
where $\phi({\bfm r})$ is the eigenfunction of the Hamiltonian~(\ref{ham}). In
QED the perturbation to the current density is due to the vacuum
polarization. The Hall current flows in the orthogonal subspace
and the corresponding correction to the vacuum polarization is given by the
first and the last terms of Eq.~(\ref{delpi}).  However, the last term of 
Eq.~(\ref{delpi}) does not vanish for $q^2=0$ and represent the change of the
photon dispersion law rather than the correction to the current density, which
is completely determined by the first Lorentz covariant term 
\begin{equation}
\delta j_I({\bfm r})=
\frac{1}{45}\frac{\alpha}{\pi} \beta^2 j_I({\bfm r})\,,
 \label{deljres}
\end{equation}
The expression for the Hall current takes the following form
\begin{equation}
I=\left(1+\frac{1}{45}\frac{\alpha}{\pi} \beta^2\right)\int
j_I({\bfm r}){\rm d}r_E{\rm d}r_B\,. \label{curres}
\end{equation}
The integral in Eq.~(\ref{curres}) can be expressed through the
derivative of the electron energy  ${\cal E}$ in $\Phi$
\begin{equation}
\int j_I({\bfm r}){\rm d}r_E{\rm d}r_B= -\frac{e}{e^\prime}\frac{{\rm
d} {\cal E}}{{\rm d}\Phi}\,, \label{int}
\end{equation}
see {\it e.g.} Ref.~\cite{Yen}. Thus our final expression for the 
Hall current reads
\begin{equation}
I=-\frac{{\rm d}{\cal E}_t}{{\rm d}\Phi}\,, \label{curfin}
\end{equation}
where ${\cal E}_t$ is the total energy of the electrons contributing
to the current. As has been shown in
Ref.~\cite{Lau1}, the flux $\Phi$ acts as a quantum pump: changing
it by $n$ quanta $\Phi_0^\prime$ results in a net transfer of $n\nu$
electrons across the ribbon, which corresponds to an energy
variation of $n\nu e^*V$. Thus for the Hall conductivity one gets
\begin{equation}
R_H^{-1}=\nu\frac{e^*}{\Phi_0^\prime}\,. \label{rkdef}
\end{equation}
Putting together
Eqs.~(\ref{estar},\ref{phistar},\ref{rkdef}) we obtain the
final expression for the von Klitzing constant
\begin{equation}
R_K^{-1}=2\alpha\left[1+\frac{2}{45}\frac{\alpha}{\pi}\beta^2\right],
\label{res}
\end{equation}
or in  physical units
\begin{equation}
R_K^{-1}=\frac{e^2}{2\pi\hbar}\left[1+\frac{2}{45}\frac{\alpha}{\pi}\left(\frac
{\hbar
eB}{c^2m^2}\right)^2\right]. \label{ressi}
\end{equation}
We would like to emphasize that the characteristic distance of the
vacuum fluctuations resulting in the correction to the Hall
conductivity is given by  the electron Compton wavelength of order
$10^{-12}~m$, which is far smaller than the actual thickness  of the
layer where  the electrons are localized, of order  $10^{-8}~m$.
Thus the correction to $R_K$ is due to an intrinsically
three-dimensional effect, which  is not prohibited by the
topological and gauge invariance arguments developed  in  two
dimensions.

The correction term in Eq.~(\ref{ressi}) can be rewritten as follows
\begin{equation}
\frac{2}{45}\frac{\alpha}{\pi}\left(\frac{B}{B_0}\right)^2,
\label{cor}
\end{equation}
where $B_0=c^2m^2/(\hbar e)\approx 4.41\cdot 10^{9}$~T. A typical
value of the magnetic field in current experiments corresponds to
$B/B_0\sim 10^{-8}$.  Thus  numerically Eq.~(\ref{cor}) amounts to a
tiny $10^{-20}$ correction. This  is well beyond the available
accuracy of the von Klitzing constant determination, which is about
ten parts per billion \cite{MohTay}.  However, this accuracy is
limited mainly by the absence of an independent standard of
resistance.  Studying the variation of $R_K$ with $B$  does not have
this restriction and can be performed by means of a different
experimental method and with a presumably significantly higher
accuracy.  A renowned
example of a similar phenomenon is given by the system of neutral
$K_S$ and $K_L$ mesons, where the absolute experimental accuracy for
the mass difference is about twelve orders of magnitude  higher than
for the average mass \cite{PDG}. Thus it is an open question whether
the evidence of  Eq.~(\ref{res}) can be  obtained with the available
experimental facilities.  

On the other hand, there is no fundamental reason which rules out the
possibility of the observation of the phenomenon in a dedicated
future experiment. A possible scheme of such an experiment involves two
identical samples assembled in a single electric circuit and exposed  to
different magnetic fields. The effect becomes observable when the Hall voltage
difference between the samples due the correction term in Eq.~(\ref{res})
reaches the resolution of the measuring device, {\it e.g.}  based on the
Josephson frequency/voltage conversion. Note that the voltage difference can be
increased by orders of magnitude if one uses stronger magnetic field and larger
values of the Hall current. We would like to emphasize that 
the  quantum Hall conductance is topologically protected against
any other type of corrections including the finite size effects
\cite{Lau1,Hal1}, which otherwise would  mask the tiny effect of vacuum
polarization.

In summary, the leading QED correction  to the  quantum mechanical
result for the Hall conductivity is derived.  It results in a weak
dependence of the universal von Klitzing constant on the magnetic
field strength. This  remarkable and unexpected manifestation of a
fine nonlinear quantum field effect in a collective phenomenon in
condensed matter merits a dedicated experimental analysis.

\begin{acknowledgements}
I am grateful to  K. Melnikov for the discussions and cross-checks. I would like
to thank  J. Beamish, A. Czarnecki and V. Rubakov for useful communication and
careful reading the manuscript.  This work is supported by the Alberta Ingenuity
foundation and NSERC.
\end{acknowledgements}



\begin{thebibliography}{99}


\bibitem{KDP} K. von Klitzing, G. Dorda, and M. Pepper, Phys. Rev. Lett. {\bf 45}, 494 (1980).

\bibitem{TSG}  D.C. Tsui, H.L. St\"ormer, and A.C. Gossard, Phys. Rev. Lett. {\bf 48}, 1559 (1982)

\bibitem{Lau2} R.B. Laughlin, Phys. Rev. Lett. {\bf 50}, 1395 (1983).

\bibitem{Hal2} B.I. Halperin, Phys. Rev. Lett. {\bf 52}, 1583 (1983).

\bibitem{MohTay} P.J. Mohr and  B.N. Taylor, Rev. Mod. Phys. {\bf 77}, 1 (2005), and references therein.

\bibitem{Lau1} R.B. Laughlin, Phys. Rev. B {\bf 23}, 5632 (1981).

\bibitem{Hal1} B.I. Halperin, Phys. Rev. B {\bf 25}, 2185 (1982).

\bibitem{TKNN} D.J. Thouless, M. Kohmoto, M.P. Nightingale, and M. den Nijs, Phys. Rev. Lett. {\bf 49}, 405 (1982).

\bibitem{Sim} B. Simon, Phys. Rev. Lett. {\bf 51}, 2167 (1983).

\bibitem{AvrSei} J.E. Avron and R. Seiler, Phys. Rev. Lett. {\bf 54}, 259 (1985).


\bibitem{Yen} D.R. Yennie,  Rev. Mod. Phys. {\bf 59}, 781 (1987).


\bibitem{CasLep} W.E. Caswell and G.P. Lepage, Phys. Lett. B {\bf 167},  437 (1986).

\bibitem{PinSot1} A. Pineda and J. Soto, Nucl. Phys. Proc. Suppl. {\bf 64},  428 (1998).

\bibitem{KniPen} B.A. Kniehl and A.A. Penin, Nucl. Phys. {\bf B563}, 200 (1999).

\bibitem{Sch} J. Schwinger, Phys. Rev.  {\bf 82}, 664 (1951).

\bibitem{Adl} S.L. Adler, Ann. Phys. (N.Y.) {\bf 67}, 599 (1971).

\bibitem{LosSko} Yu.M. Loskutov and V.V. Skobelev., Phys. Lett. A {\bf 36}, 405 (1971).

\bibitem{PDG} C. Amsler et al., Phys. Lett. B {\bf 667}, 1 (2008).

\end{thebibliography}
\end{document}